\documentclass{article}
\usepackage{graphicx}
\graphicspath{{converted_graphics/}}

\begin{document}

\begin{center}
{\huge Aerodynamic Stability and the Growth }%
\vskip1pt
{\huge of Triangular Snow Crystals }%
\vskip10pt

{\Large K. G. Libbrecht and H. M. Arnold}\vskip4pt

{Department of Physics, California Institute of Technology}\vskip-1pt

{Pasadena, California 91125}\vskip-1pt

\vskip18pt

\hrule\vskip1pt \hrule\vskip14pt
\end{center}

\medskip \noindent \textbf{Abstract}: We describe laboratory-grown snow
crystals that exhibit a triangular, plate-like morphology, and we show that
the occurrence of these crystals is much more frequent than one would expect
from random growth perturbations of the more-typical hexagonal forms. We
then describe an aerodynamic model that explains the formation of these
crystals. A single growth perturbation on one facet of a hexagonal plate
leads to air flow around the crystal that promotes the growth of alternating
facets. Aerodynamic effects thus produce a weak growth instability that can
cause hexagonal plates to develop into triangular plates. This mechanism
solves a very old puzzle, as observers have been documenting the unexplained
appearance of triangular snow crystals in nature for nearly two centuries.

\section{Introduction}

Complex patterns and structures often emerge spontaneously when crystals
grow, yielding a great variety of faceted, branched, and other forms. This
is readily seen, for example, in the well-known morphological diversity
found in naturally occurring mineral crystals \cite{minerals}. Suppressing
structure formation is often desired when growing large commercial crystals,
but exploiting the phenomenon provides a possible route for using nanoscale
self-assembly as a manufacturing tool \cite{nano}. Whether the goal is to
reduce, enhance, or otherwise control, there has been considerable interest
from a number of fronts in characterizing and understanding the detailed
physical mechanisms that produce ordered structures from disordered
precursors during solidification \cite{solidification, shapes}.

The molecular dynamics involved in the condensation of disordered molecules
into a regular crystalline lattice is remarkably complex, involving a number
of many-body effects over different length scales and time scales \cite%
{saito}. As a result, calculating dynamical properties like crystal growth
rates from first principles is generally not yet possible. Predicting growth
morphologies has also proven quite difficult, with few overarching theories
connecting the many disparate physical mechanisms that govern growth
behaviors.

A well-known and oft-studied example of structure formation during crystal
growth is the formation of ice crystals from water vapor. In the atmosphere
these are called snow crystals, and they fall from the clouds with a
remarkable diversity of morphologies, including simple plate-like and
columnar forms, elaborately branched plates, hollow columns, capped columns,
and many others \cite{fieldguide}. We have been studying the detailed
physics of snow crystals as a case study in crystal growth, with the hope
that developing a comprehensive mechanistic model for this specific system
will shed light on the more general problem of structure formation during
solidification \cite{libbrechtreview}.

\begin{figure}[t] 
  \centering
  \includegraphics[width=4.1in,keepaspectratio]{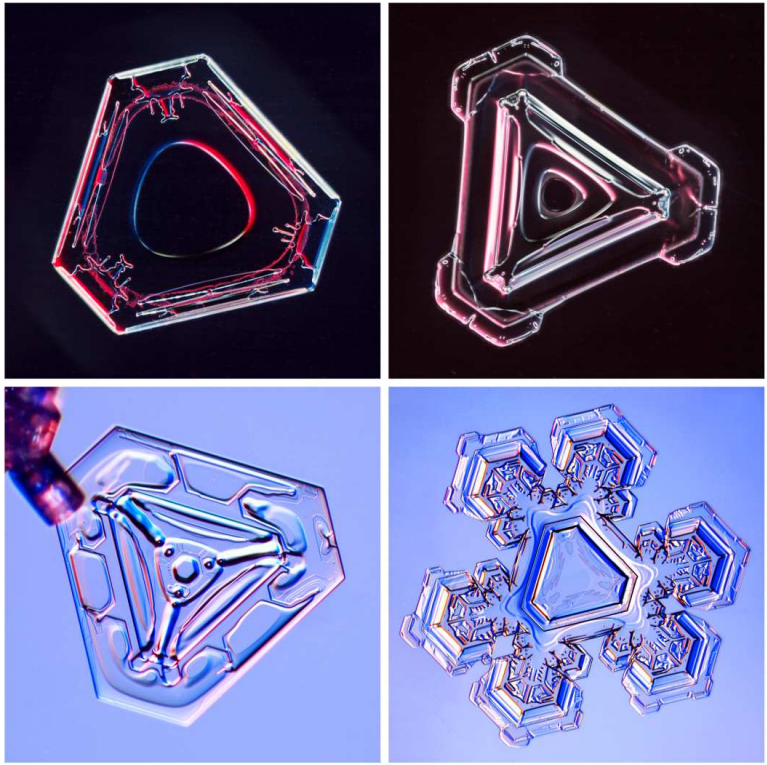}
  \caption{Examples of natural snow
crystals exhibiting triangular morphologies (from \protect\cite{snowflakes}%
). The equivalent diameters (defined by $d=(4A/\protect\pi )^{1/2}$, where $A
$ is the projected 2D crystal area) range from 1 to 3 mm. The lower right
example shows a crystal with an initial truncated triangular morphology
(outlined by the central surface markings) that subsequently grew plate-like
branches.}
  \label{triangles2x}
\end{figure}

Within the menagerie of known snow crystal morphologies, observers have long
documented the occurrence of forms exhibiting a peculiar three-fold
symmetry, as shown in Figure 1. Scoresby reported triangular forms as early
as 1820 \cite{scoresby}, and Bentley and Humphreys presented several dozen
examples with triangular morphologies in their well-known 1931 compilation
of snow crystal photographs \cite{bentley}. Although triangular crystals are
usually small and relatively rare, they are visually distinctive and fairly
easy to find in nature. While the usual six-fold symmetry seen in snow
crystals arises from the intrinsic hexagonal symmetry of the ice crystal
lattice, to our knowledge the formation of triangular shapes has never
before been explained.

In this paper we first present experimental measurements of laboratory-grown
ice crystals showing that triangular plate-like crystals do in fact form
more frequently than would be expected with random growth perturbations.
From these data, as well as observations of triangular forms in nature, we
conclude that some physical mechanism is required that coordinates the rapid
growth of alternating crystal facets. Following this, we then describe an
aerodynamic model that explains the growth and stability of triangular snow
crystals.

\section{Laboratory-Grown Crystals}

To examine the phenomenon of triangular snow crystals under controlled
conditions, we used a convection growth chamber \cite{chamber} to produce
thin ice crystal plates grown in air at a pressure of one bar. Simple,
plate-like snow crystals grown at low supersaturation typically exhibit
hexagonal morphologies, but we observed small numbers of crystals with
three-fold symmetry at temperatures near -2C and between -10C and -15C \cite%
{data2008}. Triangular crystals were especially prevalent when grown at a
temperature of -10C in low water vapor supersaturations, so all the data
shown below were taken at -10C with a supersaturation of 1.4 percent. Under
these conditions, approximately five percent of the plate-like crystals we
observed had a truncated triangular appearance, and examples with nearly
perfect equilateral-triangle morphologies were readily found.

We first looked at the prevalence of triangular and other non-hexagonal
morphologies by nucleating crystals in our chamber and recording video
images of an unbiased sampling of all plate-like crystals that fell on a
substrate at the bottom of the chamber after 5-10 minutes of growth in free
fall. We ignored non-plate-like, blocky forms, which are present in
relatively small numbers at this temperature \cite{data2008}.

\begin{figure}[t] 
  \centering
  \includegraphics[width=4.5in,height=3.6in,keepaspectratio]{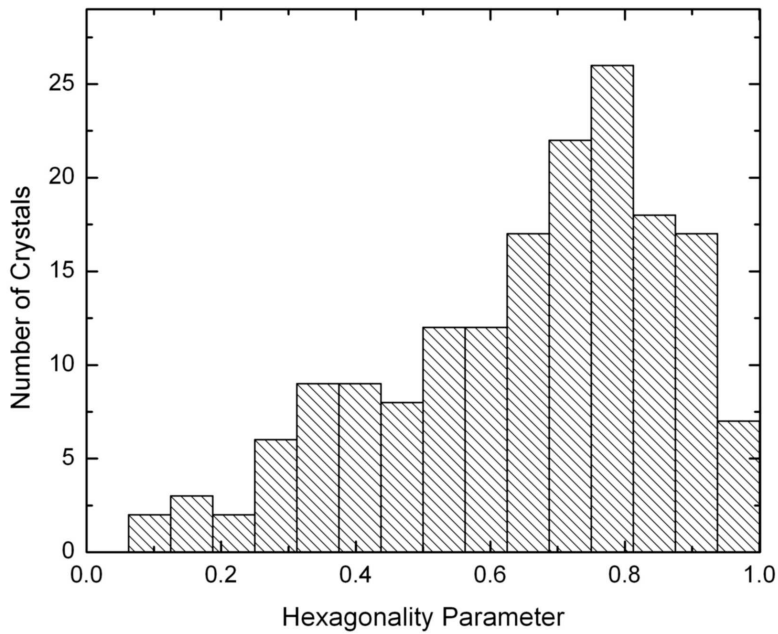}
  \caption{Distribution of laboratory-grown crystals as a
function of the hexagonality parameter $H$ defined in the text. These data
are from an unbiased sample of plate-like crystals grown at -10 C with a
water vapor supersaturation of 1.4 percent.}
  \label{fig:GraphHrevisedx}
\end{figure}

\begin{figure}[ht] 
  \centering
  \includegraphics[width=4.5in,height=3.56in,keepaspectratio]{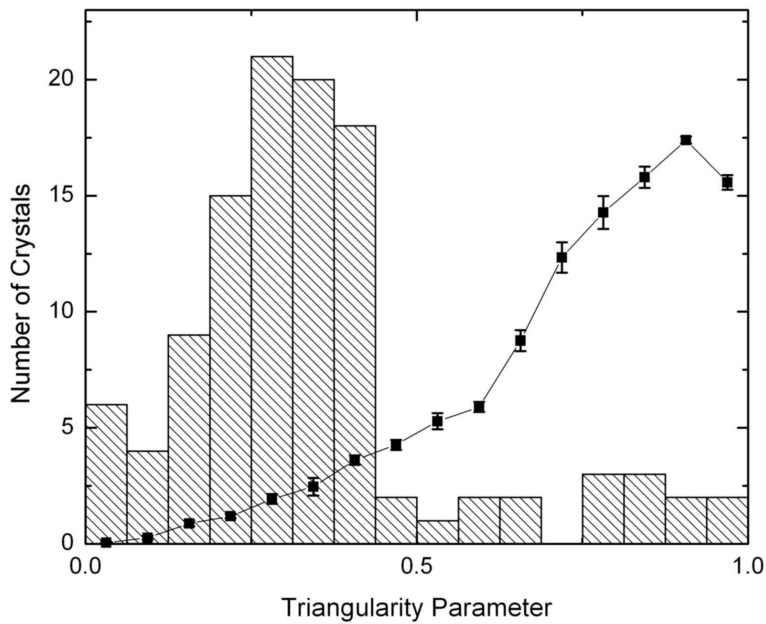}
  \caption{Distribution of laboratory grown crystals with $H<1/3$, as a function
of the triangularity parameter $T$ defined in the text (bars). The line
shows a Monte Carlo model for $T$, described in the text, which assumes
random growth perturbations of a hexagonal plate. Error bars show the
uncertainty in the model, estimated by varying a number of details in the
calculations. The data and model clearly show that crystals with a
triangular morphology (small $T$) are much more common than one would expect
from random growth perturbations.}
  \label{fig:GraphTrevisedx}
\end{figure}

We defined a \textquotedblleft hexagonality\textquotedblright\ parameter $%
H=L_{1}/L_{6}$ for the crystal plates, where $L_{1}$ and $L_{6}$ are the
lengths of the shortest and longest prism facets, respectively. $H$ is close
to unity when a plate is nearly hexagonal, while $H$ is smaller for any of a
variety of odd-shaped plates. Figure 2 shows the measured $H$ distribution
for an unbiased sample from our data. Crystals with $H>3/4$ appeared nearly
hexagonal to the eye, and this plot shows that the crystals in our sample
were mostly hexagonal with relatively small distortions. We say a plate-like
crystal has an \textquotedblleft extreme\textquotedblright\ morphology (far
from hexagonal) if $H<1/3$.

From the subset of extreme crystals, we then measured a \textquotedblleft
triangularity\textquotedblright\ parameter $T=L_{3}/L_{4}$, where $L_{3}$
and $L_{4}$ are the lengths of the third and fourth smallest facets,
respectively. This parameter was small if, and only if, the morphology was
that of a truncated triangle, and we observed $T\rightarrow 0$ if, and only
if, the morphology was nearly that of an equilateral triangle. Figure 3
shows the $T$ distribution for extreme crystals only. Most of the extreme
plate crystals were clearly in the shape of truncated triangles to the eye,
and this is reflected in the fact that the $T$ distribution is skewed to
lower values. Figure 4 shows a sample of some of the extreme crystals we
observed in our data.

\begin{figure}[p] 
  \centering
  \includegraphics[width=4.5in,height=6.06in,keepaspectratio]{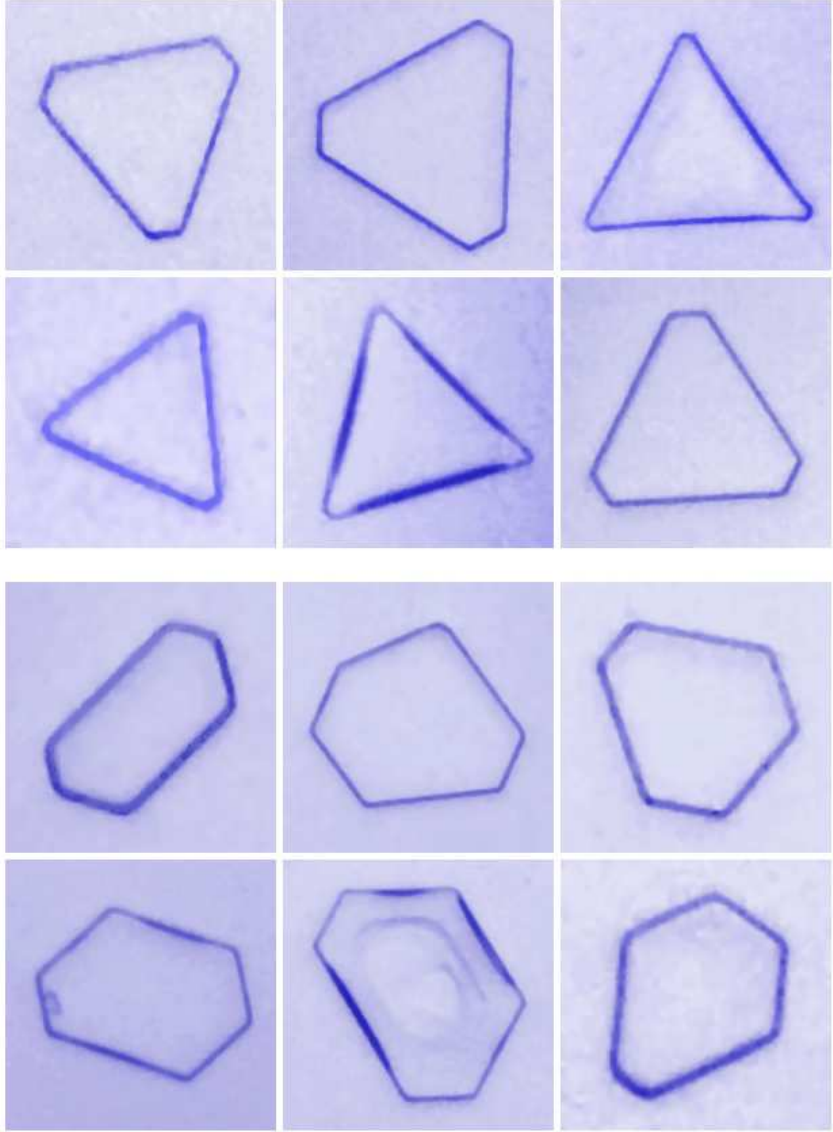}
  \caption{Examples
of some extreme ($H < 1/3$) crystal morphologies seen in our
data. The top six images show crystals with $T < 1/2$, while the
bottom six show crystals with $T > 1/2$. The former were much
more common than the latter, as can be seen from Figure 3. Equivalent
diameters ranged from roughly 50 to 100 microns.}
  \label{fig:combined3}
\end{figure}

We used a simple Monte Carlo simulation to investigate the null hypothesis
that our observations were due entirely to random variations in the growth
rates of the different prism facets. We generated crystals for which the
perpendicular growth velocity of each facet was constant and chosen from the
same random distribution. We then \textquotedblleft grew\textquotedblright\
these crystals, selected ones with $H < 1/3$, and from many such
crystals we generated a T distribution to compare with our data. The results
are shown in Figure 3. Here the error bars in the model were estimated by
using a variety of sensible random distribution functions (for example,
truncated normal distributions with different means and standard deviations)
to generate crystal growth rates. In our simulations, random fluctuations in
the growth rates of the six facets produced a variety of odd shapes,
including trapezoidal, diamond-shaped, and other forms. The simulations did
not show a preponderance of triangular crystals over other non-hexagonal
shapes, in stark contrast to our data.

These considerations provide convincing evidence that, at least under some
growth conditions, triangular morphologies are substantially more abundant
than one would expect from random fluctuations in the growth of hexagonal
plates. From this we conclude that some non-random mechanism is responsible
for the growth of snow crystal plates with three-fold symmetry. In
particular, this mechanism must somehow coordinate the growth of the facets
so that they alternate between slow and fast growth around the crystal.

\begin{figure}[p] 
  \centering
  \includegraphics[width=4.5in,height=6.66in,keepaspectratio]{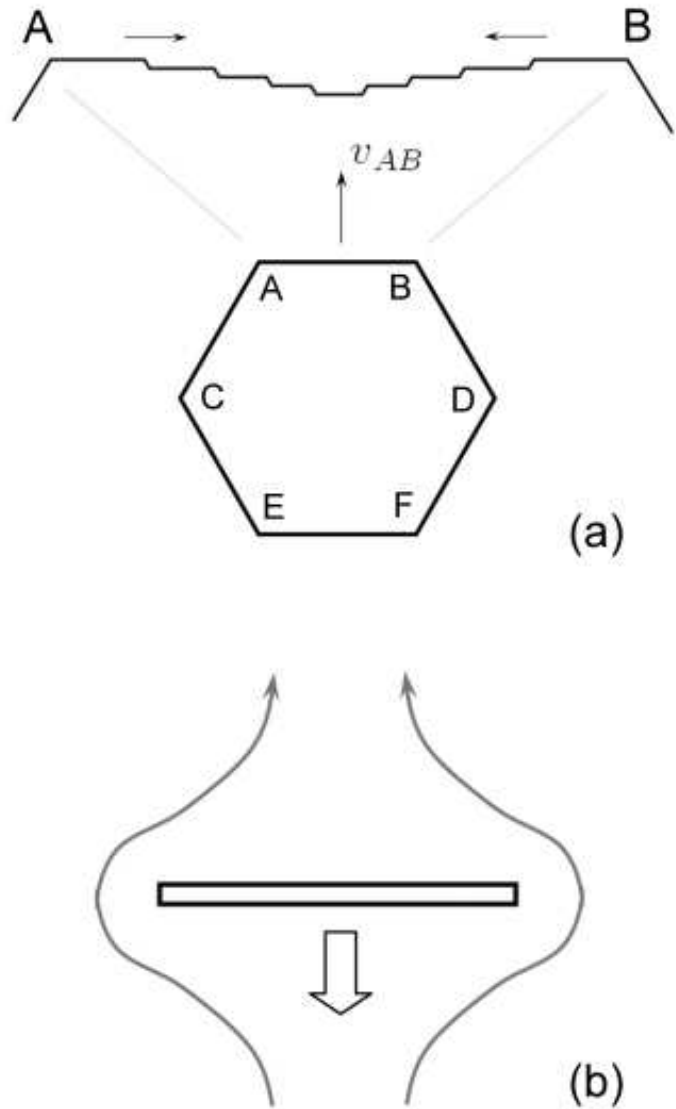}
  \caption{(a) The formation of facets in
a simple hexagonal plate-like crystal, as described in the text. The top
figure shows a close-up of the $AB$ facet, exaggerated to show molecular
steps on the surface. (b) A schematic depiction of airflow around a falling
hexagonal plate crystal (seen from the side). }
  \label{fig:hex1ax}
\end{figure}

\section{An Aerodynamic Model}

To examine the effects of aerodynamics on snow crystal morphology, first
consider the case of a thin hexagonal plate crystal, as shown in Figure 5.
The crystal has six prism facets, and each grows outward at some
perpendicular growth velocity (in Figure 5, for example, $v_{AB}$ is the
growth velocity of the $AB$ facet). For a symmetrical crystal, all six prism
facets are the same length and all six growth velocities are equal.

The growth of a faceted crystal is limited partially by water vapor
diffusion through the surrounding air and partially by attachment kinetics
at the crystal surface. The two effects together result in facet surfaces
that are slightly concave at the molecular level, as shown in Figure 5,
although they may appear perfectly flat optically \cite{saito}. Nucleation
of new molecular terraces occurs near the corners (points $A$ and $B$ in
Figure 5), where the supersaturation is highest. The molecular steps then
propagate inward, traveling more slowly near the facet centers where the
supersaturation is lower. The combined effects of surface attachment
kinetics and diffusion-limited growth thus automatically establish the
concave shape of each facet surface.

The perpendicular growth velocity $v_{AB}$ of the $AB$ facet is primarily
determined by the nucleation rates near points $A$ and $B$. In the
symmetrical case, the rates at $A$ and $B$ are equal, giving the picture
shown in Figure 5. If the nucleation rate were slightly greater at $A$, then
the picture would be distorted and the facet surface would be tilted
slightly relative to the ice lattice. If the nucleation rate at $A$ were
substantially greater than at $B$, then terraces generated at $A$ would
propagate all the way across a vicinal surface from $A$ to $B$ \cite{saito}.
In general, we see that $v_{AB}$ is determined by the greater of the two
nucleation rates at $A$ and $B$.

As a growing crystal falls, air resistance causes the basal faces to become
oriented perpendicular to the fall velocity (i.e., parallel to the ground),
as shown in Figure 5. Atmospheric halo observations have shown that thin
plate crystals with diameters $>50$ $\mu $m can orient to within
a few degrees of horizontal under ideal conditions \cite{tape}. Smaller
crystals have such low terminal velocities that aerodynamic forces orient
them relative to the local wind velocity rather than orienting them relative
to the ground.

Air flowing around a falling crystal tends to increase its growth via the
well-known ventilation effect \cite{vent, fukuta99}. The airflow produces an
effective increase in supersaturation where the edges stick out farthest and
the resulting flow is the fastest (see Figure 5). In the case of a falling
plate, this aerodynamic effect mostly increases the growth of the thin edges
of the plate (i.e., the prism facets).

\begin{figure}[t] 
  \centering
  \includegraphics[width=4in,keepaspectratio]{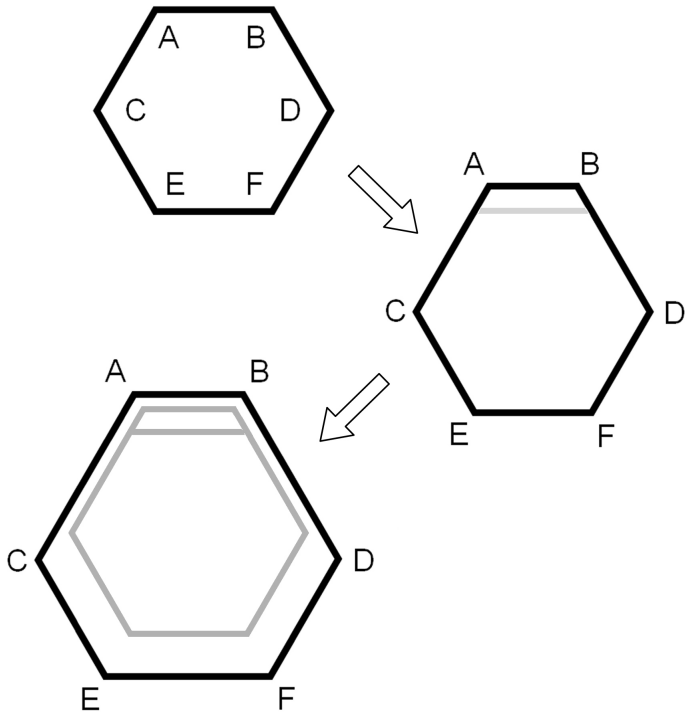}
  \caption{An aerodynamic model for the
formation of triangular crystals, as described in the text.}
  \label{fig:hex2}
\end{figure}

With this overall picture in mind, we now consider the case shown in Figure
6. Here we start with a hexagonal plate and assume a small growth
perturbation somewhere on the $AB$ facet that breaks the six-fold symmetry
and makes $v_{AB}$ greater than the other five facet growth rates. This
perturbation could come from a crystal dislocation, a step-generating
chemical impurity on the surface, a piece of dust, or perhaps some other
mechanism. Regardless of the origin of the initial symmetry-breaking
perturbation, the larger growth rate $v_{AB}$ initially results in the
distorted crystal shape shown as the second stage in Figure 6.

Once this initial asymmetry appears, the additional material produces an
increased aerodynamic drag on one side of the plate. This in turn changes
the orientation of the falling crystal in such a way that the points $E$ and 
$F$ tip downward, toward the fall direction. Relative to the original
horizontal orientation, air flowing around the tilted crystal now increases
the supersaturation at points $E$ and $F$ while decreasing it at $A$ and $B$%
. As a result, the nucleation rates at points $E$ and $F$ increase.

The growth rate of a facet is determined mainly by the greater of the
nucleation rates at its two corners (from the discussion of Figure 5 above),
and for our tilted crystal we see that the nucleation rates at $E$ and $F$
must be greater than at $C$ and $D$. Thus we must have that $v_{CE}\approx
v_{EF}\approx v_{DF}$ to a rough approximation, and furthermore these
velocities must all be greater than $v_{AC}$ and $v_{BD}$ (while $v_{AB}$ is
somewhat ill-determined throughout because of the assumed growth
perturbation). After a period of additional growth in these conditions,
shown in the last stage of Figure 6, we find that the length of facet $EF$
has increased relative to $CE$ and $DF$. This means that $AB$, $CE$, and $DF$
are now the three shortest facets, so the crystal has begun to assume a
slight triangular shape.

Once aerodynamics produces the shape shown in the last stage of Figure 6,
two effects combine to accentuate the triangular morphology. First, the
shorter facets ($AB$, $CE$, and $DF$ in this case) stick out farther into
the supersaturated air, so the Mullins-Sekerka instability \cite{saito}
tends to increase $v_{AB}$, $v_{CE}$, and $v_{DF}$ relative to the other
three facets. An important aspect of this instability is that it takes less
mass to grow a surface of reduced size, so overall mass flow considerations
in diffusion-limited growth tend to increase the growth of the shorter
facets. Second, airflow is faster around the shorter facets, because they
stick out farther, and this also increases their growth relative to the
longer facets via the ventilation effect.

The end result is that an initial symmetry-breaking perturbation results in
a growth morphology that becomes more triangular with time. Only one initial
perturbation is necessary, and no coordination intrinsic to the molecular
structure of the crystal need be present. The coordination of the growth
rates of alternating facets is initiated by the aerodynamics of the falling
crystal.

An interesting feature of our model is that a triangular plate morphology is
both aerodynamically stable and stable against additional growth
perturbations. Once a plate takes the form of an equilateral triangle $%
(T\rightarrow 0)$, subsequent growth perturbations cannot change the
morphology, as long as the plate remains faceted. This is not true for
hexagonal plates, so even small initial perturbations would eventually
result in triangular shapes. This triangular growth instability is weak,
however, so it would not be a dominant effect under typical conditions.

The tendency to form triangular crystals is enhanced if the growth rate
increases rapidly with increasing supersaturation, as this accentuates the
growth differences resulting from aerodynamic effects. This may be why our
observations at -10C and low supersaturation yielded an especially high
fraction of triangular crystals.

\section{Conclusions}

Observations of natural snow crystals have long suggested that triangular
morphologies are more common than one would expect from random growth
perturbations, and that triangular forms are generally more common than
other non-hexagonal plates. This conclusion is strongly reinforced by our
measurements of laboratory-grown crystals. Since a trigonal symmetry is not
present in the underlying crystalline lattice, some external mechanism is
required to explain the formation of these crystals.

We believe that the aerodynamic model described above explains the
occurrence of snow crystal plates with triangular morphologies, both in the
lab and in nature. In particular, this mechanism qualitatively shows how a
single symmetry-breaking perturbation in a hexagonal plate can result in the
growth of a triangular morphology.

Unfortunately, producing a more complete, quantitative aerodynamic model
will be difficult. Researchers have only recently developed viable numerical
techniques for modeling the diffusion-limited growth of faceted crystals 
\cite{gg}, even for relatively simple physical cases. Adding aerodynamic
instabilities and their resulting growth changes in full 3D will likely be a
considerable challenge. Nevertheless, even our simple qualitative model
makes a number of testable predictions.

As we suggested above, for example, triangular morphologies are more likely
to appear when the attachment coefficient is a strong function of
supersaturation. Additional measurements at other temperatures and
supersaturations would help confirm this behavior. More directly, \textit{in
situ} observations of electrodynamically levitated crystals \cite{electro}
could be used to examine the transition from hexagonal to triangular forms
in real time, imposing different airflows around the crystal to test many
aspects of our aerodynamic model in detail.

Finally, we hope that an improved understanding of the mechanisms
controlling ice growth may shed light on other systems, whenever growth is
affected by diffusion-limited transport, surface attachment kinetics, and
particle transport via large-scale flows. The growth of triangular snow
crystals is another piece in the puzzle that describes the many
interconnected mechanisms by which complex structures emerge spontaneously
during solidification.


\begin{thebibliography}{99}
\bibitem{minerals} Prinz, M., Harlow, G., and Peters, J., \textquotedblleft
Simon \& Schuster's Guide to Rocks \& Minerals,\textquotedblright\ Simon \&
Schuster (New York), 1978.

\bibitem{nano} Imai, H., \textquotedblleft Self-organized formation of
hierarchical structures,\textquotedblright\ Springer-Verlag (Berlin), 2007.

\bibitem{solidification} Kolasinski, K. W., \textquotedblleft Solid
structure formation during the liquid/solid phase
transition,\textquotedblright\ Curr. Opin. Solid State \& Mater. Sci., 11,
76-85, 2007.

\bibitem{shapes} Ben-Jacob, E., and Garik, P., \textquotedblleft Ordered
shapes in nonequilibrium growth,\textquotedblright\ Physica D, 38, 16-28,
1989.

\bibitem{saito} Saito, Y., \textquotedblleft Statistical physics of crystal
growth,\textquotedblright\ World Scientific (Singapore), 1996.

\bibitem{fieldguide} Libbrecht, K. G., \textquotedblleft Ken Libbrecht's
Field Guide to Snowflakes,\textquotedblright\ Voyageur Press (St. Paul),
2006.

\bibitem{libbrechtreview} Libbrecht, K. G., \textquotedblleft The physics of
snow crystals,\textquotedblright\ Rep. Prog. Phys., 68, 855-895, 2005.

\bibitem{snowflakes} Libbrecht, K. G., \textquotedblleft
Snowflakes,\textquotedblright\ Voyageur Press (St. Paul), 2008.

\bibitem{scoresby} Scoresby, W. \textquotedblleft An Account of the Arctic
Regions with a History and Description of the Northern
Whale-Fishery,\textquotedblright\ Archibald Constable Publishing (London),
1820.

\bibitem{bentley} Bentley, W. A., and Humphreys, W. J., \textquotedblleft
Snow Crystals,\textquotedblright\ McGraw-Hill (New York) 1931.

\bibitem{chamber} Libbrecht, K. G., and Morrison, H. C., \textquotedblleft A
Convection Chamber for Measuring Ice Crystal Growth
Dynamics,\textquotedblright\ arXiv:0809.4869, 2008.

\bibitem{data2008} Libbrecht, K. G., Morrison, H. C., and Faber, B.,
\textquotedblleft Measurements of Snow Crystal Growth Dynamics in a
Free-fall Convection Chamber,\textquotedblright\ arXiv:0811.2994, 2008.

\bibitem{tape} Tape, W., \textquotedblleft Atmospheric
Halos,\textquotedblright\ American Geophysical Union, 1994.

\bibitem{vent} Keller, V. W., Hallett, J. \textquotedblleft Influence of air
velocity on the habit of ice crystal growth from the
vapor,\textquotedblright\ J. Cryst. Growth 60, 91-106, 1982.

\bibitem{fukuta99} Fukuta, N. and Takahashi, T., \textquotedblleft The
growth of atmospheric ice crystals: A summary of findings in vertical
supercooled cloud tunnel studies,\textquotedblright\ J. Atmos. Sci. 56,
1963-79 (1999).

\bibitem{gg} Gravner, J., and Griffeath, D., \textquotedblleft Modeling
snow-crystal growth: A three-dimensional mesoscopic
approach,\textquotedblright\ Phys. Rev. E, 79, 011601, 2009.

\bibitem{electro} Swanson, B. D., Bacon, M. J.; Davis, E. J., et al.,
\textquotedblleft Electrodynamic trapping and manipulation of ice
crystals,\textquotedblright\ Quart. J. Roy. Meteor. Soc., 125, 1039-1058,
1999.
\end{thebibliography}
\end{document}